# Combined effects of particle geometry and applied vibrations on the mechanics and strength of entangled materials

Saeed Pezeshki, and Francois Barthelat*

Laboratory for Advanced Materials and Bioinspiration, Department of Mechanical Engineering, University of Colorado, 427 UCB, 1111 Engineering Dr, Boulder, CO 80309, USA

* Corresponding author: francois.barthelat@colorado.edu

**Abstract**

Entangled materials offer attractive structural features including tensile strength and large deformations, combined with infinite assembly and disassembly capabilities. How the geometry of individual particles governs entanglement, and in turn translates into macroscopic structural properties, provides a rich landscape in terms of mechanics and offers intriguing possibilities in terms of structural design. Despite this potential and recent report on these materials, there are major knowledge gaps on the entanglement mechanisms and how they can generate strength. In particular, vibrations are known to have strong effects on entanglement and disentanglement but the exact mechanisms underlying these observations are unknown. In this report we present tensile tests and discrete element method (DEM) simulations on bundles of entangled staple-like particles that capture the combined effects of particle geometry and vibrations on local entanglement, tensile force chains and strength. We show that standard steel staples with $\theta=90°$ crown-leg angle initially entangle better than $\theta=20°$ modified staples because of their more "open" geometry. However, as vibrations are applied entanglement increase faster in $\theta=20°$ bundles, so that they develop strong and stable tensile force chains, producing bundles which are almost ten times stronger than $\theta=90°$ bundles. Both tensile strength and entanglement density increase with vibrations and also with deformations, up to a steady state value. At that point the rate of entanglement equals the rate of disentanglement, and each of these rates remains relatively high. Finally, we show that vibration can be used as a manipulation strategy to either entangle or disentangle staple-like entangled granular materials, with confinement playing a significant role in determining whether vibration promotes entanglement or disentanglement. This work provides a fundamental understanding of how particle geometry and vibrations can be used to manipulate the properties of entangled materials, which can lead to better design guidelines for lightweight, reversible materials and structures and aggregate architectures.

**I. Introduction:**

Traditional granular materials are composed of spherical or convex particles that lack intrinsic tensile strength [1] unless interstitial fluid [2] or adhesives [3–5] are used to maintain structural integrity. In contrast, entangled materials based on non-convex particles with hook-like or bard-like features exhibit unique mechanical properties driven by interlocking geometry and dynamic internal rearrangements. "Hexapods" made of slender "legs" with high aspect ratios jam and entangle, forming free-standing structures [6], offering intriguing possibilities in terms of structural design and architecture [7,8], such as aleatory architecture [9]. An exciting approach in designing the entangled granular materials is to tailor the geometry of individual grains towards more "extreme" geometries that induce better entanglement. Extreme particle geometries such as star-like particles [10], or particles with branches with hooks and barbs (e.g. U-shaped [11–15], or Z-shaped staple-like particles[16]) can provide better entanglement and higher tensile strength. Prior works have demonstrated that geometric parameters—such as the leg length [11], or crown-leg angle ($\theta$) [17,18] in U-shaped particles—govern the entanglement, and consequently, the strength of staple bundles. Understanding how the geometry of individual particles govern entanglement, and in turn translate into strength provides a rich landscape in terms of mechanics and design. As part of this effort, various experimental approaches have been used to assess entanglement strength. Perhaps the simplest of these experiments consists of measuring how many particles can be pulled from a bundle against gravity [16,18–21]. More complex experiments have measured angle of repose [22,23], or the stability of long free-standing columns [24,25] and short columns subjected to vibrations [11]. Other mechanical tests on entangled materials have included tensile tests [12,17], compressive tests [26] and flexural tests[16]. While these experiments measure "macroscopic" mechanical properties for bundles of entangled particles, they provide limited insights into the fundamental mechanisms of entanglement and disentanglement at the local level, and little information on how force chains appear and evolve with deformation. Such insights can be gained using Monte Carlo simulations[18] or Discrete element models[14,17,21]. However, a thorough understanding on the formation of entanglement and force chains at short and longer range in a bundle is still missing. In particular, the effects of mechanical vibrations on entanglement are not well known. Previous studies show that vibration strongly influences the structure and dynamics of granular matter based on convex grains. These studies show that vibration can improve the packing density (e.g., rods[27], spherical particles[28], cubical particle[29], platonic solids[30]). Rods, as anisotropic granular materials, show clear ordering under vertical vibration: they begin in a disordered, low-packing-fraction state and gradually organize into a dense nematic state, with their long axes aligning along the container walls[27,31]. When vibrations are applied, the packing fraction of spherical grains increases until a steady state, although fluctuations persist[32]. In entangled granular materials, vibrations can also change entanglement density which plays a more important role than packing density in determining their structural property[11,16,19–21]. For example, a recent study used a network-based approach to consider how the entanglement network of C-shaped particles evolves with vibration, revealing a percolation threshold [20]. U-shaped "smarticles" and worm collectives also show that oscillations can either tighten or loosen entanglement, letting the material switch between stiff and fluid-like states[33]. Vibrations can also be used to disassemble entangled materials: Pillars made of entangled staple-like particles progressive collapse when subjected to vibrations, at a rate which can be used as a measure of entanglement and strength[11]. The aim of this study was to shed more light on the combined effects of particle geometry and applied vibrations on entanglement density, force chains and tensile strength, using tensile experiments and discrete element models on staple-like particles.

## II. Experiments

The main objective of the experiments was to measure the combined effects of staple geometry and applied vibrations on the tensile response of bundles of staples. We used standard steel office staples (Swingline, IL) with the dimensions shown on Fig. 1a. Preparing bundles of staples was relatively simple: Sticks of staples were immersed in an acetone bath to dissolve the weak adhesive holding the staples together, and the detached staples were then rinsed and stored in containers (Fig. 1b). Our previous studies have shown that changing the angle between the crown and the legs (the crown-leg angle $\theta$) has strong effects on entanglement and on the strength of the bundle[17]. For this study we used staples with a standard crown leg angles of $\theta=90°$, and a smaller crown leg angles $\theta=20°$ which we created by pressing sticks of staples against a 3D printed tool (Fig. 1c) before separating the staples.

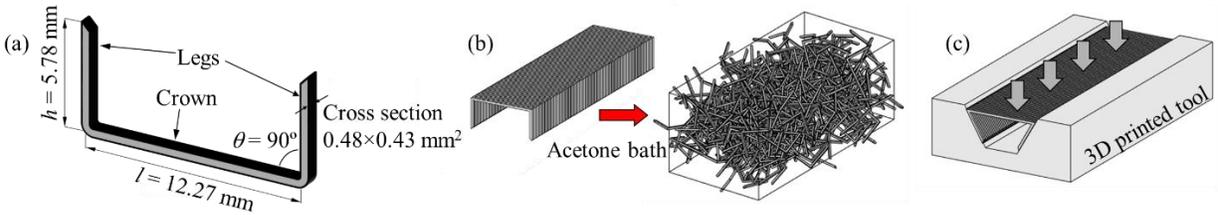

**Figure 1:** (a) Dimensions of an individual staple with a standard crown-leg angle $\theta=90°$; (b) An acetone bath disassembles a "stick" of staples into individual staples; (c) On some batches the crown-leg angle was reduced to $\theta=20°$ using a custom 3D printed tool.

For each experiment, 1,000 staples were pluviated into a 60 × 40 × 30 mm³ acrylic box. The box was then placed on a vertical vibration stand (Eisco Labs Vibration Generator) and subjected to vertical sinusoidal vibrations with an amplitude of 2.5 mm and a frequency of 30 Hz. This particular combination produced visible changes of conformation in the volume of bundle, displacing staples within the volume of the bundle. The main parameter we varied to change the "amount" of vibration delivered to the bundle was the number of vibration cycles $N$. After pluviation and vibration, the bundle was transferred onto a horizontal Teflon surface within a horizontal tensile machine (ADMET expert 4000 Micro Tester). Custom grippers made of six vertical nails each were used to transmit pulling displacements to the ends of the bundle [17]. The tensile tests were performed in displacement-controlled conditions at pulling rate of 10 mm/min, corresponding a strain rate of ~ $4\times10^{-3}$/s.

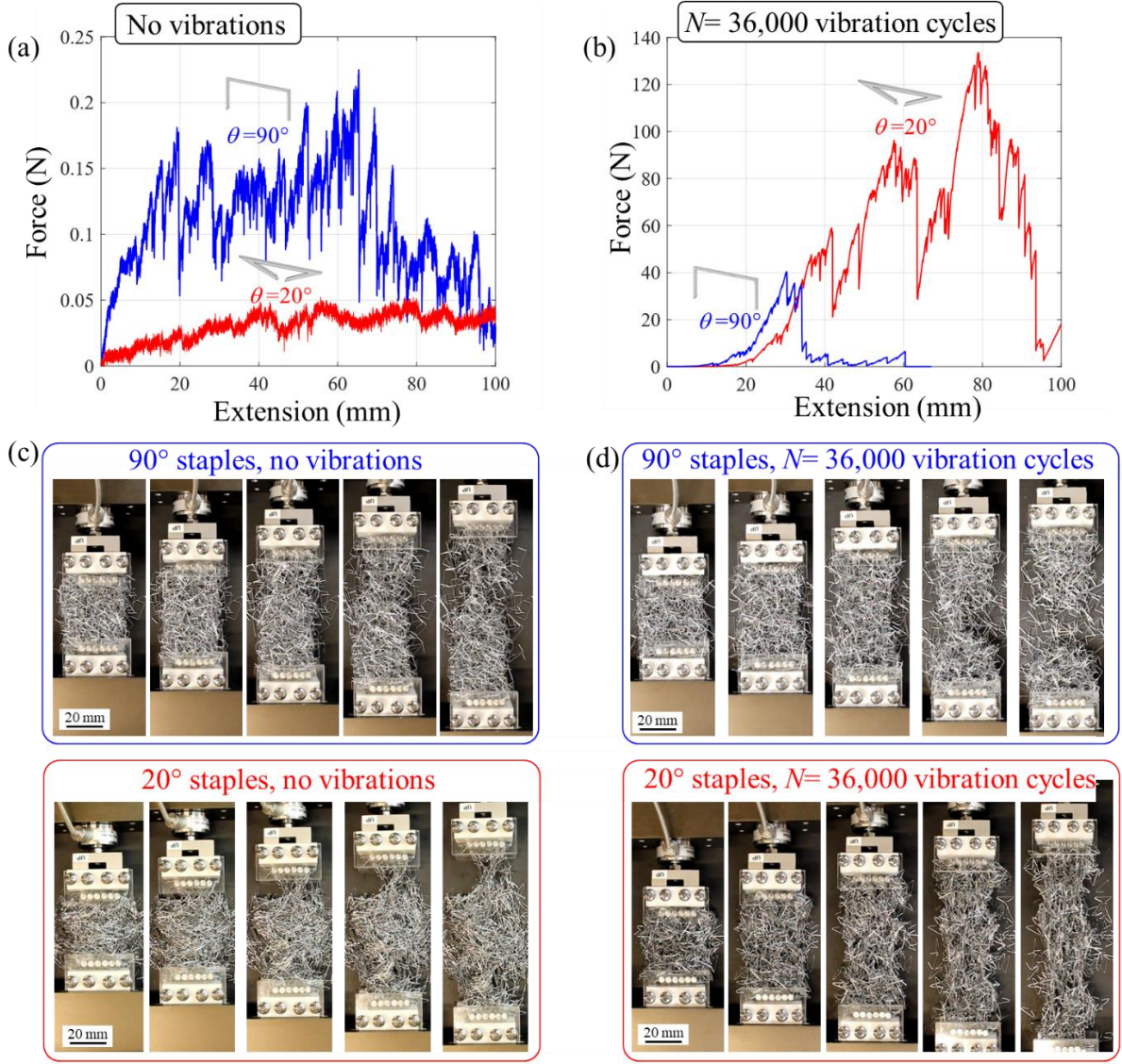

**Figure 2:** Tensile tests of staple-like entangled bundles: (a) Typical tensile force-extension curves for staple-like entangled granular material with $\theta$ =90° and $\theta$ =20° crown-leg angles prepared with no vibration (*N*=0), and (b) subjected to *N*= 36,000 cycles of vibration; (c-d) Images acquired during the tensile tests.

Fig. 2 shows typical tensile test results for bundles of $\theta$=90° and $\theta$=20° staples subjected to no vibration (Fig. 2a) and subjected to *N*=36,000 cycles of vibrations (Fig. 2b). All force-extension curves shared the same broad characteristics: relatively large extensions and jagged appearances with numerous force peaks and valleys, which we attributed to tensile force chains breaking and reforming dynamically within the deforming bundle [17]. The representative results of Fig. 2 also show the strong combined effects of staple geometry and applied vibrations on the strength and deformability of the bundle. With no applied vibrations (Fig. 2a) the bundles are relatively weak, with the $\theta=90°$ bundles being about four times stronger than the $\theta=20°$ bundles. When $N=36,000$ vibration cycles were applied prior to the tensile test, the $\theta=90°$ bundles were 200 times stronger than with no vibrations applied (but the bundles failed at smaller deformations) and the $\theta=20°$ bundles were about 2000 times stronger than with no vibrations applied (Fig. 2b). Interestingly, with no vibration the $\theta=90°$ bundles are therefore the strongest, while with vibration applied it is the $\theta=20°$ bundles that are the strongest. This results clearly shows that vibration increases entanglement within the staple bundles, with effects modulated by the geometry of the staples. Fig. 2c shows images from the tensile tests, showing a relatively homogeneous deformation pattern in $\theta=90°$ bundles. In contrast, the $\theta=20°$ bundles subjected to no vibration cycle formed long but loosely connected chains, suggesting limited entanglement which resulted in reduced strength. Bundles of $\theta=20°$ bundles subjected to 36,000 vibration cycles developed stronger, more interconnected chains which could sustain tensile strains in excess of 100%. Failure was generally preceded by a local thinning in the density of staples (similar to necking in metals), with the exception of the vibrated $\theta=20°$ bundle: In that case the neck that initially formed resisted deformation and propagated in the entire bundle, suggesting that the neck region is stronger than the rest of the bundle (in a mechanisms similar to tensile "drawing" in glassy polymers [34]).

We now turn our attention to measuring the tensile strength of the bundles subjected to various amounts of vibrations, and to developing adequate statistical approaches to compare the combined effects of staple geometry and applied vibration. We performed five tensile tests for each configuration considered in this study, each producing one datapoint for strength. However, as pointed above, a typical force-extension tests consists of multiple failure events corresponding to the failure of individual force chains. Therefore, instead of measuring a single value for strength for each tensile test, we considered each local peak force (local maximum) that occurred during the test [12]. For each curve, a simple algorithm was used to detect local maxima and minima, which we defined as points where the force suddenly dropped by more than 10% of its value (Fig 3a, c). Using this method, each tensile test produced between 100 and 3,000 individual values for bundle strength. This collection of strengths was then plotted as cumulative distribution functions (CDFs) for $\theta=90°$ bundles subjected to $N=0$ (Fig. 3b) and for $N=36,000$ cycles of vibrations (Fig. 3d). As expected, the distribution is relatively broad, with tensile strengths ranging from about 0.015 to 0.324 N for $N=0$ vibrations. We then fitted these distributions with a variety of existing statistical models. Following Franklin [12], we first fitted the experimental CDFs with Weibull distributions using a least square approach, which produced acceptable results with a coefficient of determination $R^2 = 0.9296$ for no vibration case, and $R^2 = 0.9572$ for 36,000 cycles of vibration (Fig. 3b, d). We found however that other statistical models could produce better fitting results. Fig. 3b, d shows fits from normal, log-normal and exponential distributions. Among these models, a comparison of $R^2$ values revealed that the log-normal distribution produced the best fit of our experimental data.

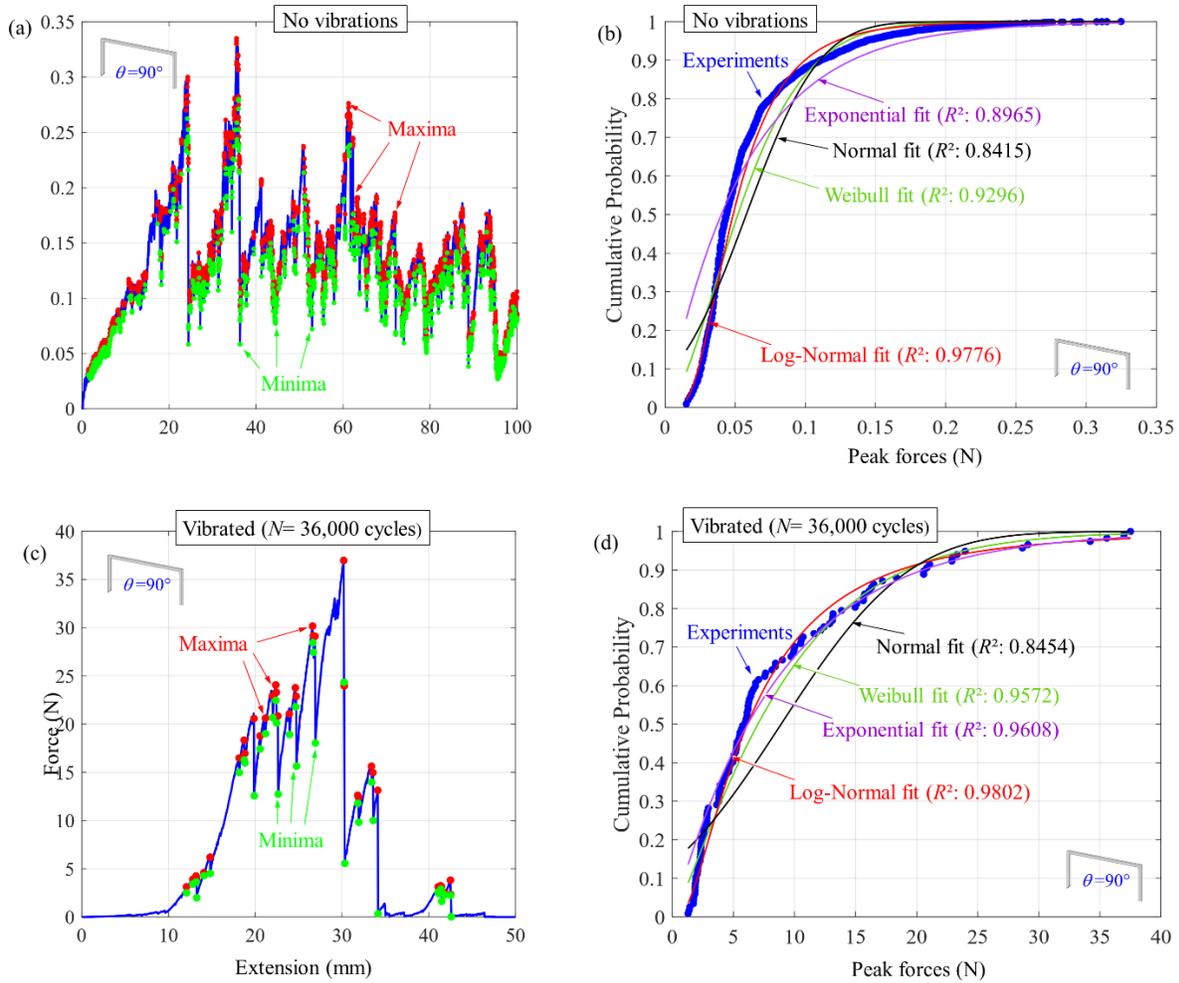

**Figure 3**: (a) Typical force-extension curve for a $\theta=90°$ bundle (no vibrations applied), showing local maxima and minima; (b) Cumulative distribution function (CDF) of peak forces fitted with various statistical models for the $\theta=90°$ bundle (no vibrations applied); (c) Force-extension curve for a $\theta=90°$ bundle with $N=36,000$ vibration cycles; (d) corresponding CDF of peak forces fitted with various statistical models.

The lognormal distribution typically captures failure mechanisms based on multiplicative degradation processes [35–37], which is consistent with bundles of staples where failure is the result of many small, independent random perturbations or local "shocks" (staples slipping, staples rotations, staples sudden ejection, plastic deformation of staples). The CDF of the log-normal distribution is written:

$$P(F_s) = \frac{1}{2}\left[1 + erf\left(\frac{\ln F_s - \langle \ln F_s \rangle}{\sqrt{2}\sigma}\right)\right] \quad (1)$$

Where *erf* is the Gauss error function, $F_s$ is the bundle strength and $\sigma$ is the standard deviation of $\ln F_s$. The corresponding probability density function (PDF) is:

$$p(F_s) = \frac{1}{F_s \sigma \sqrt{2\pi}} \exp\left(\frac{-(\ln F_s - \langle \ln F_s \rangle)^2}{2\sigma^2}\right) \quad (2)$$

For each configuration, we therefore characterized the strength of the bundle by two parameters: The average of the natural logarithm of the strength $\langle \ln F_s \rangle$, and the breadth of the distribution $\sigma$.

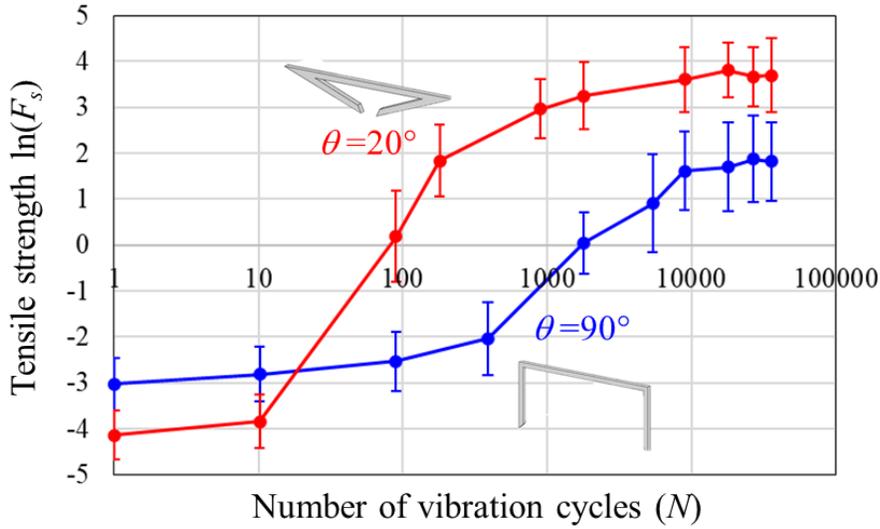

**Figure 4:** Tensile strength measured as $\ln(F_s)$ and plotted as function of the number of vibrations cycles $N$ applied prior to the tensile test on $\theta=90°$ bundles (blue) and $\theta=20°$ bundles (red).

Fig. 4 shows the strength of the bundle varies as function of applied vibration cycles for $\theta=20°$ and $\theta=90°$ bundles. For both staple geometries the tensile strength increases with the number of applied vibrations cycles, until the strength saturates at about $N=10,000$ cycles. As mentioned above, the $\theta=20°$ bundles are weaker than the $\theta=90°$ bundles when no vibrations are applied.

However, the strength of the $\theta=20°$ bundles increases faster than the $\theta=90°$ bundle with vibrations, a crossover occurring between $N=10$ and $100$ cycles, and a saturation average strength for the $\theta=20°$ bundles about eight times higher than the $\theta=90°$ (at saturation $\langle \ln F_{s20} \rangle \sim \langle \ln F_{s90} \rangle + 2$, so that $\langle F_{s20} \rangle \sim 8 \langle F_{s90} \rangle$). We finally note that the breadths of the strength distribution (length of the error bars on Fig. 4) are remarkably consistent across all combinations of vibrations and staple angles. The increase in strength with higher number of applied vibration cycles is most likely the result of increased density of entanglement in the bundle. This question, as well as others related to the effect of geometry and entanglement dynamics during deformation, are explored in the next section.

### III. Discrete Element Models (DEM)

The discrete element method, initially developed for traditional granular materials [38,39], is naturally well suited for bundles of discrete staples. We recently developed a DEM model[17] for staples using the granular package in LAMMPS[40]. The dimensions of the staples were identical to the staples used for experiments, with a crown length $l = 12.27$ mm, and leg lengths $w = 5.78$ mm. Individual staples were discretized with spheres (diameter $d = 0.45$ mm, consistent with the thickness of the backbone). Individual spheres on the backbone were spaced by a distance $s=0.45$ mm and connected by flexible elastic bonds using the Bonded Particle Model [41,42] (Fig. 5). The elastic elements could experience combined bending (stiffness $k_b=EI/s$ with $I$ =second moment of area), axial deformation (stiffness $k_r=AE/s$ with $A$ =cross sectional area), torsion (stiffness $k_t=JG/s$ with $G$ =shear modulus and $J$ =polar moment of area) and transverse shear (stiffness $k_s=AG/s$). Using the actual modulus of steel produced bonds with such high stiffness that the system became numerically unstable, especially for the vibration as a dynamic process, unless extremely small timesteps were used which makes simulations for the larger 1000-staple systems prohibitive computationally. We therefore used the same approach as our previous paper on modeling of bundle of staples where we scaled the elastic properties of the bonds (as well as other parameters, such as mass density and gravity)[17]. This approach is also consistent with using a "secant" modulus for the material of the staples (lower than the modulus of steel) to approximate the elastic-plastic response of steel [43–45]. Spheres belonging to different staples interacted via Hertzian contact with a linear history-dependent friction model[46–48]. We used a value of $\mu = 0.3$ for static and dynamic friction coefficients (in our previous study we shows that $\mu$ has minor effects on pair strength compared with geometrical interlocking[17]).

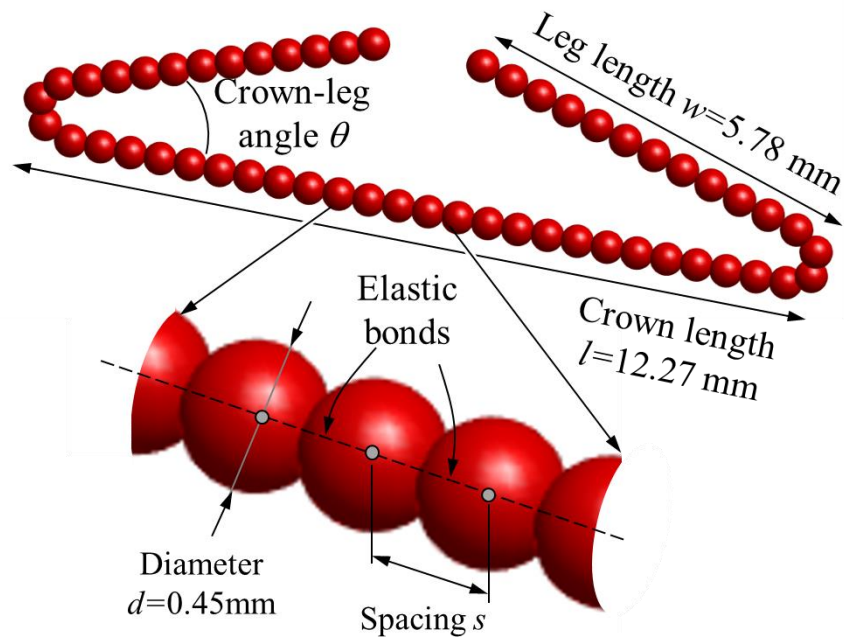

**Figure 5:** DEM model of an individual staple based on discrete spheres joined with elastic bonds.

The simulations started with pluviating 1,000 staples into a 60 × 40 × 30 mm³ container under gravity (Fig. 6a). For the simulations, we set the mass of the individual staples $m$ and gravity $g$ so that $El^2 / mg = 10^8$, which resulted in a response similar to the experiments during both the pluviation and vibration processes (no excessive rebounds and no excessive deformations of the individual staples from gravity only). The interactions between the walls of the box and the staples were modeled using the Hertz contact formulation, with the same parameters as those applied to staple–staple contacts. The container was then subjected to vertical oscillations to simulate vibrations, with an amplitude identical to the experiments. Because of time scaling, the frequency used in the simulation was adjusted so the bundle responded to vibrations in a way identical to the experiments: In both cases staples underwent motion, but not as much as to eject the staples from the container.

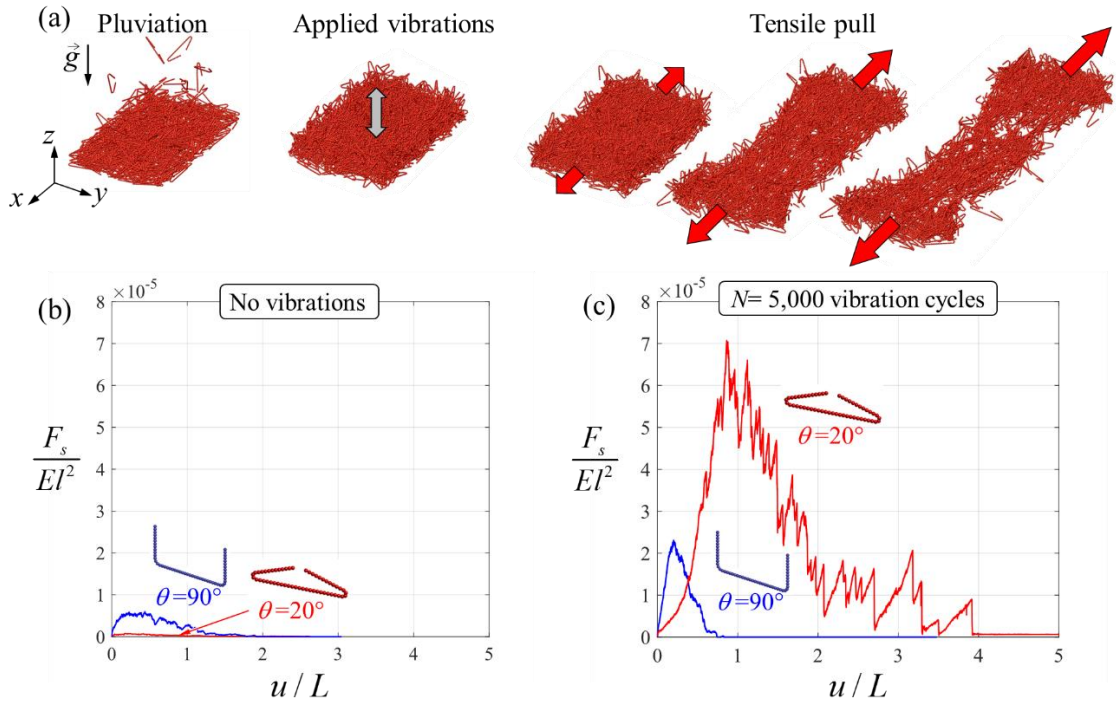

**Figure 6:** DEM simulations of tensile tests on entangled bundles: (a) The three main steps for the simulations: Pluviation, applied vibrations, and tensile pull; Force-extension curves for $\theta=20°$ (red) and $\theta=90°$ (blue) bundles prepared with (b) no vibrations ($N=0$) and (c) $N=5,000$ vibration cycles.

Once the vibration step was completed with the desired number of cycles applied, the system was allowed to relax, and six vertical rods were created at each end of the bundle to duplicate the experimental clamping conditions. These rods interacted with the staples using the same contact law as the staple-wall interactions. The sudden insertion of the rods introduced localized disturbances, so the system was again allowed to fully relax. Following this step the side walls of the container were removed, while the bottom wall and gravity were retained to maintain consistent loading conditions. The two sets of boundary rods were then moved apart at a slow rate to impose a quasi-static tensile loading on the bundle (Fig. 6a). The displacement rate was chosen to ensure negligible inertial effects, allowing the system to remain in a state of global quasi-static equilibrium throughout the deformation process. Fig. 6 b,c show typical force-extension curves from these simulations for $\theta=20°$ and $\theta=90°$ bundles, and for $N=0$ and $N=5,000$ applied vibration cycles. The results predict the same trends as the experiments (Fig. 2a,b): With no vibrations ($N=0$) $\theta=90°$ bundles are stronger than $\theta=20°$ bundles, and when $N=5,000$ vibrations are applied both types of bundles get much stronger, the $\theta=20°$ bundles becoming stronger than the $\theta=90°$ bundles. To complete this dataset, we applied different amounts of vibrations on the $\theta=20°$ and $\theta=90°$ bundles, simulating three realizations for each combination. Using the force extension curves, we then collected local peak forces using the same protocol as for the experiments. Fig. 7 shows the cumulative density function for strength for the 90° bundles subjected to no vibrations, and subjected to 5,000 vibration cycles, with several theoretical distributions fitted to these results. Interestingly, and in consistence with the experiments, it is again the log-normal distribution which produced the best fit on the simulation results.

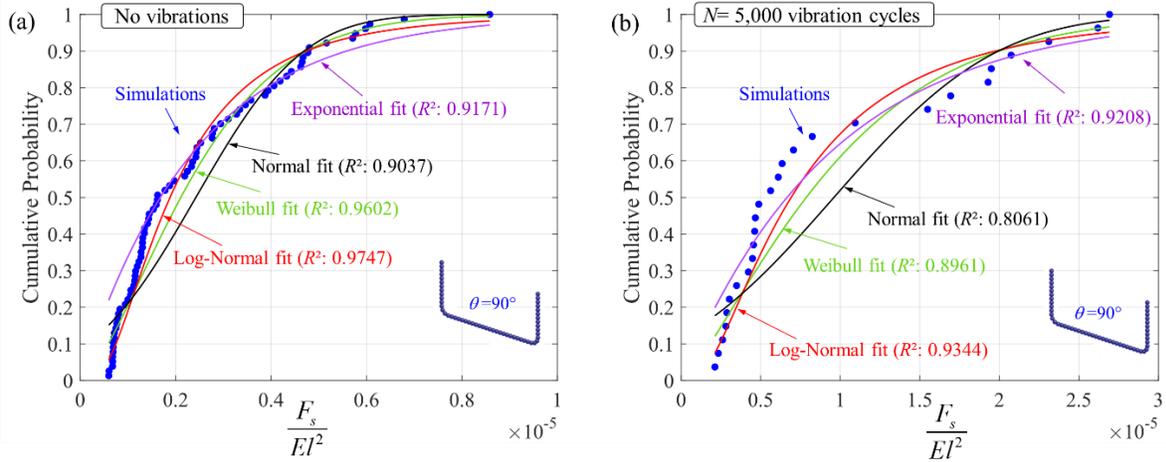

**Figure 7:** Cumulative distribution function (CDF) of peak forces from $\theta=90°$ bundles fitted with various statistical models: (a) no vibration applied and (b) $N$=5,000 vibration cycles applied.

Using this data, we extracted the log-normal parameters strength $<ln\,(F_s/El^2)>$ and $\sigma$ (standard deviation of $ln\,(F_s/El^2)$) for each configuration. The results (Fig. 8) show that $\theta=20°$ bundles are weaker than $\theta=90°$ bundles for no or few vibrations, but that the 20° bundles get stronger from about $N\sim100$ cycles of applied vibrations, both designs reaching a steady state strength at about $N\sim300$ cycles. These trends (strength increases with crossover between $\theta=20°$ and $\theta=90°$ bundles, steady state strength) are remarkably consistent with the experimental results of Fig. 4. However, in the model this sequence occurs at smaller numbers of cycles compared to the experiments. This discrepancy can be due the limitation of BPM package which we modeled staples with elastic bonds, where in the experiments some of the staples undergo plastic deformations. In addition, the dissipation during the vibration process maybe larger in the experiments than in the simulations (especially against the acrylic container). Nevertheless, the DEM simulations captured the experimental trends well and we did not attempt to further refine the model. The main use we made of the DEM were to collect detailed information on mechanisms at the level of individual staples, including the formation of tensile force chains, and the evolution of entanglement with vibrations and deformations.

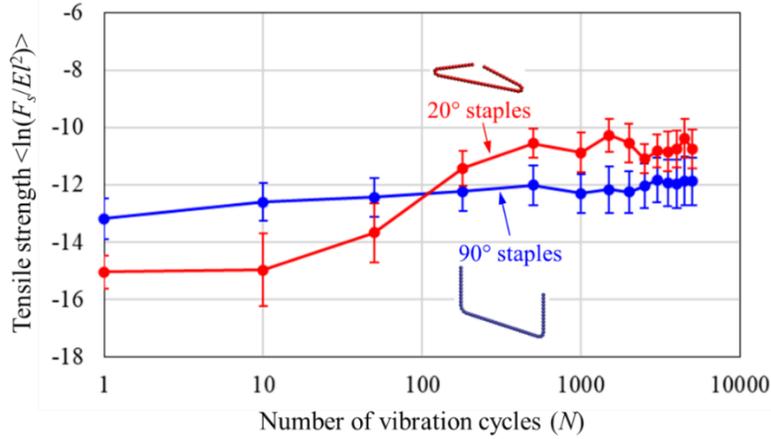

**Figure 8:** Normalized bundle strengths as function of number of applied vibration cycles for $\theta=20°$ bundles (red) and $\theta=90°$ bundles (blue).

To visualize these force chains, we first collected the magnitude of the forces carried through the backbone of individual staples. We then identified the group of staples that carried most of the applied tensile force. Fig. 9 shows the staples that carried 75% of the total load, colored according to the force they transfer. All other staples, which carry only the remaining 25% of the load, are shown in light gray.

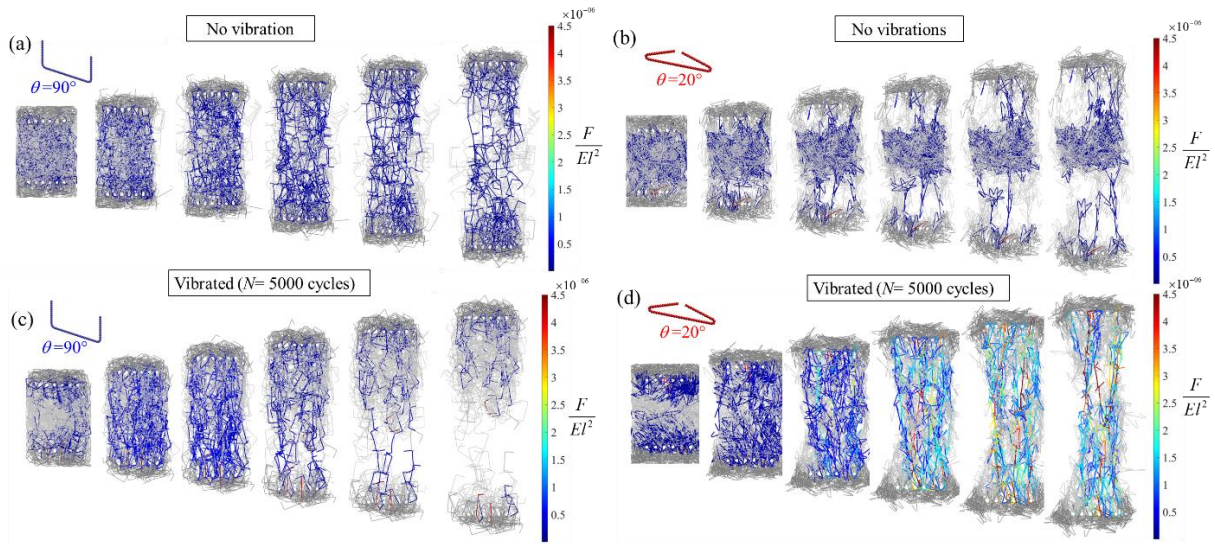

**Figure 9:** Snapshots of force chains from DEM simulations subjected to no vibration (a): $\theta=90°$ bundles and (b): $\theta=20°$ bundles, and subjected to 5,000 cycles of vibration (c): $\theta=90°$ bundles and (d): $\theta=20°$ bundles.

These snapshots of Fig. 9 a,b reveal that when the bundles are not vibrated, the tensile forces in the $\theta=90°$ bundles are distributed more homogenously compared to the $\theta=20°$ bundles, where force chains are shorter and localization occurs earlier (localization also occurs at two different place, in consistence with the experiments). As a result, the 90° bundle shows better tensile performance compared to the 20° bundle (when no vibrations are applied). Fig. 9c-d show that when samples are prepared with 5,000 vibration cycles, the network of force chains that develop in tension are denser than with no vibrations. Vibrations seem to be particularly effective for the 20° bundles where necking occurs in tension, but where staples located within the neck region carry more forces than the rest of bundle so that the neck region is mechanically stable and propagates in entire bundle (in consistence with the experiments). Vibrations therefore promote the generation of force chains in bundles in tension, which can be explained by the amount of entanglement that develops during the vibration preparation process, and also during the tensile test itself. Following a protocol used in our previous report [17], we tracked the entanglement of individual staples by defining 'nets' for each staple—flat regions partially enclosed by the crown and legs (Fig. 10a)—that can engage and "catch" other staples in the bundle (Fig. 10b). We then counted, using a three-dimensional intersection algorithm, the number $T$ of neighboring staples that intersected any of the two nets of each staple in the bundle, also requiring that the central staple crossed its neighbors though one of the neighbor's nets ("reciprocal" engagement). Fig. 10c shows that the average entanglement $<T>$ increases as the number of applied vibration cycles is increased, the system being the most sensitive to vibrations in the $N=100$ to 1000 range. The results also show that the entanglement in the $\theta=20°$ bundles are consistently lower than the $\theta=90°$ bundles, which is consistent with our previous study on the effect of geometry [18]: $\theta=20°$ staples have a geometry which is less "open" to engaging with neighboring staples compared to $\theta=90°$ staples. However, $\theta=20°$ bundles are more sensitive to vibrations: Applying $N=5,000$ cycles to $\theta=20°$ bundles increase their average entanglement by a factor of 6, while $\theta=90°$ bundles entanglement only increases by a factor of ~2.6. The entanglements bonds between $\theta=20°$ staples are also about 10 times stronger than $\theta=90°$ staples [17]. This superior staple-staple bond strength in the $\theta=20°$ bundles, combined with their faster entanglement with vibrations, contribute to their superior tensile strength when vibrations are applied. We also tracked the average entanglement that develop in the bundles *during* tensile deformations. Fig. 10c shows that for non-vibrated bundles $<T>$ is initially relatively low, but that it increases monotonically with deformation, suggesting that idle staples are being recruited for entanglement during the deformation process. For vibrated bundles with $N=5,000$ cycles the effect is more subtle with non-monotonic trends, but with $<T>$ converging to a relatively high steady state value (about 2.7 entanglements per staples for both $\theta=20°$ and $\theta=90°$ bundles).

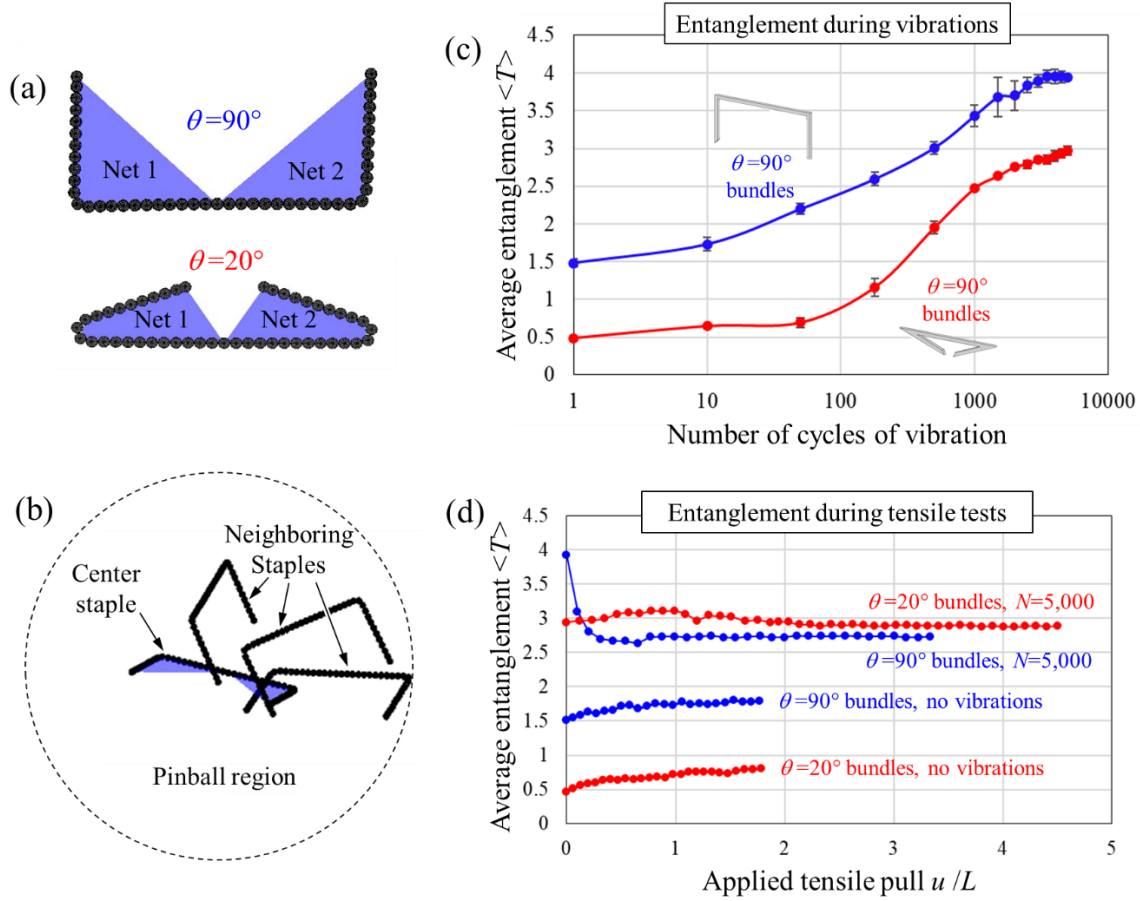

**Figure 10:** Measuring entanglement: (a) For each staple "nets" are defined as plane regions which can (b) engage with neighboring staples; (c) average entanglement in the bundle as function of applied vibration cycles; (d) average entanglement as function of tensile pull.

The results of Fig. 10 clearly show that the average entanglement $<T>$ reaches a state value when a sufficient number of vibration cycles, or a sufficient tensile pull, are applied. In this final section of the analysis, we examine possible fluctuations in entanglement when these steady state regimes are reached. A first important observation was that in both experiments and models, local rearrangements of staples could be observed even at steady state. To measure this effect, we computed entanglement and disentanglement rates for each staple and computed the average rates in bundles. Fig. 11a shows the average rate of entanglement, the average rate of disentanglement and net entanglement rate as function of applied vibrations. The net entanglement rate is positive as the average entanglement increases with vibrations and up to about 3000 applied cycles, at which point a steady state is reached with a net rate of entanglement of zero. However, Fig. 11a also shows that this net zero rate results from a balance between entanglement and disentanglement rates, each of these remaining significant in the bundle. This important result shows that at steady state, staple entanglements between staples still break and reform dynamically, but that the average entanglement rate in the bundle equals the disentanglement rate. Fig. 11a also shows that this activity is slightly higher in $\theta=90°$ bundles compared to $\theta=20°$ bundles (i.e. the entanglement and disentanglement rates are slightly higher), which could be explained by the more "open" structure of the $\theta=90°$ staples which favors more frequent transitions between entangled and disentangled states. Fig. 11b shows the average entanglement rates as function of applied tensile deformation. Here again, the results show that the net entanglement rates result from a balance between relatively high positive entanglement rates and disentanglement rates, which remain significant even in the steady state (beyond $u/L$~2.5). Interestingly, the entanglement and disentanglement rates are much higher in the $\theta=20°$ bundles compared to $\theta=90°$ bundles in the $u/L$~1 to 2.5 range. The $\theta=20°$ staples may be more favorable to dynamic entanglement in this range of deformation, possibly because of their capacity to recruit idle staples in the bundles. On the other hand, disentanglement is also higher, possibly because the tensile forces in $\theta=20°$ bundles are significantly higher than in $\theta=20°$ bundles, especially beyond $u/L$~1 (Fig. 6c).

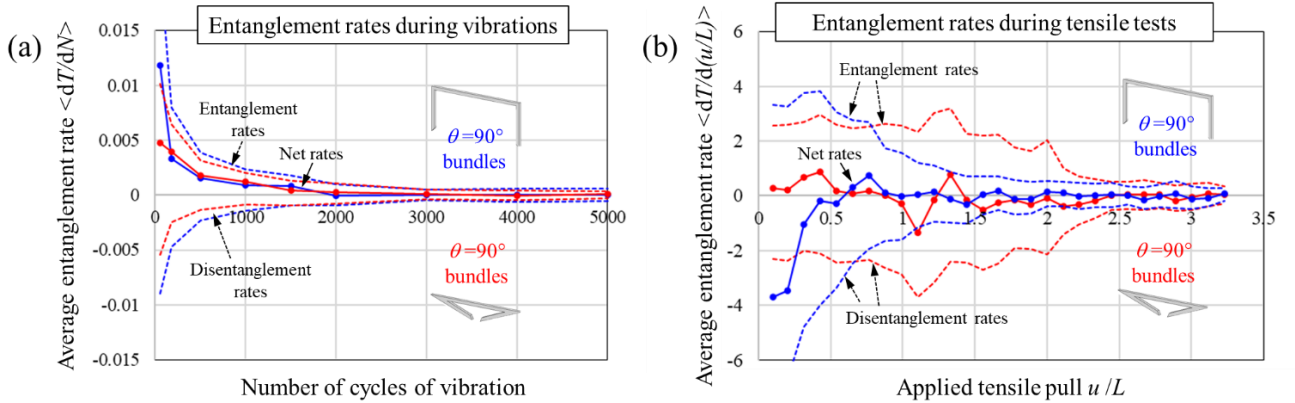

**Figure 11:** Entanglement rates, disentanglement rates and net rates for: $\theta=90°$ and $\theta=20°$ bundles as function of (a) applied cycles of vibrations and (b) applied tensile strains on bundles (that were subjected to $N=5,000$ cycles prior to pulling).

**IV. Vibration-driven disassembly**

The simulation results above clearly show that when vibrations are applied to a bundle of staple, a steady state of entanglement is reached when the entanglement rate equals the disentanglement rate. We hypothesize that this state is only possible if the bundle of staples is mechanically confined, so that even when individual staples bounce against each other locally, they remain in a dense environment of staples amenable to entanglement. In this section we show a last set of experiments where we remove some of this confinement. We first prepared dense entangled bundles with 1,000 $\theta=90°$ staples, using 36,000 cycles of vertical vibration in a container with a frequency of 30 Hz and an amplitude of 2.5mm. This entangled bundle was then transferred onto a horizontal plate with no side walls and mounted on a vibration generator. Fig. 12 shows snapshots of the bundle taken at increasing number of cycles (with frequency of 20 Hz, and amplitude of 2 mm). With no side confinements, the staples located on the side surfaces separated from when vibrations were applied. As more vibrations were applied this process continued, until the entangled bundle completely "melted" away into a thin layer of staples on the floor of the container. This result has two important implications: (1) the equilibrium between entanglement and disentanglement rates can be manipulated with confinement (In this experiment the disentanglement rate clearly outpaced the entangled rate) and (2) controlled combinations of confinement and vibrations can be applied to assemble, disassemble or change the conformation of entangled bundles.

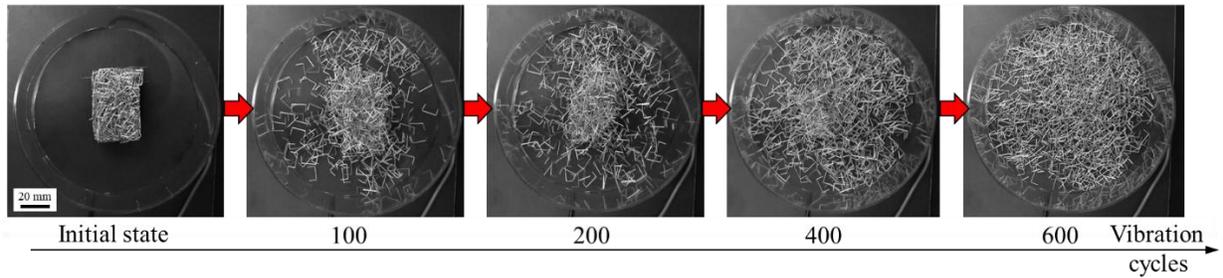

**Figure 12:** Disassembly of a block of 1,000 $\theta=90°$ staples using unconfined, vertical vibrations.

**V. Summary**

In this study we used experiments and DEM models to explore the combined effects of staple geometry and vibrations on entanglement, force transmission and on tensile strength in entangled bundles of staples. Our study reveals nontrivial and unexpected effects, which are summarized below:

- Tensile experiments reveal large deformations and jagged force-displacement responses. The strength of the bundles is a strong function of the geometry of the staples and of the amount of applied vibration prior to the tensile tests.
- Discrete element models based on discretization of the individual staples with spheres and elastic bonds capture all the features observed experimentally.
- The distributions of tensile strength in entangled bundles are broad, and in both experiments and simulations they are best captured by log-normal distributions. This result suggests that the failure of entangled bundles is governed by multiplicative degradation processes and stochastic rearrangements rather than by weakest link statistics (Weibull distribution).

- $\theta=90°$ staples have a structure with is relatively open geometrically, and overall their entanglement density is higher than in $\theta=20°$ staples which have a more "closed" geometry less amenable to engaging with neighboring staples. As a result, with little or no vibrations applied $\theta=90°$ bundles are stronger than $\theta=20°$ bundles.

- As vibrations are applied, entanglement densities increase in $\theta=90°$ bundles and $\theta=20°$ bundles. However, entanglement increases faster in $\theta=20°$ bundles, and the entanglement bond that are formed are stronger. As a result, vibrated $\theta=20°$ bundles are almost 10 times stronger than $\theta=90°$ bundles.

- Both tensile strength and average entanglement density increase with vibrations up to a steady state value. However, this does not mean that these systems become static: In these steady states the rate of entanglement equals the rate of disentanglement, each of these rates remaining relatively high.

- Entanglement densities also vary with tensile deformations. At large deformations a steady state is reached for the average entanglement, but locally the bundle remains very active as staples keep entangling and disentangling, and as force chains break and reform.

- Vibration can be used as a manipulation strategy to either entangle or disentangle staple-like entangled granular materials, with confinement playing a significant role in determining whether vibration promotes entanglement or disentanglement.

This study provides new insights into how the mechanics of staple-like entangled granular materials can be manipulated, highlighting the interplay between the geometry of the particles, vertical vibrations and confinement. A better understanding of these parameters will enable pathways to engineer granular materials that combine desirable properties such as high strength and high toughness, enabling new structural applications and a new paradigm for the assembly and disassembly of fully recyclable materials and structures. The assembly process uses mechanical stimuli only a process which is scalable and reversible. No binder is required so particles can be rapidly assembled in an infinite number of shapes, which can be reconfigured with appropriate mechanical stimuli, making them attractive for lightweight and reversible materials and structures and aggregate architectures. The fundamental understanding of entanglement gained in this project is also relevant to other physical systems, including biological structures (bird nests[49], root networks[50]), "living" entangled matter (fire ant rafts[51], worm blobs[52]), robotic materials[21] as well as colloidal assemblies[53].

**Acknowledgments**

This work was supported by the US National Science Foundation (Mechanics and Materials and Structures CMMI-2033991 and CMMI-2517927). DEM simulations were performed on the Alpine high performance computing resource at the University of Colorado Boulder. Alpine is jointly funded by the University of Colorado Boulder, the University of Colorado Anschutz, Colorado State University, and the National Science Foundation (Award 2201538).